\documentclass[11pt]{article}
\usepackage[a4paper]{geometry}
\usepackage{clic2017}
\usepackage{times}
\usepackage{latexsym}
\usepackage{graphicx}
\usepackage[colorinlistoftodos]{todonotes}
\usepackage[colorlinks=true, allcolors=blue]{hyperref}
\usepackage{footmisc}
\usepackage{multirow}

\usepackage{booktabs}

\title{Hateminers : Detecting Hate speech against Women}

\author{Punyajoy Saha\textsuperscript{$1$}, Binny Mathew\textsuperscript{$2$},  Pawan Goyal\textsuperscript{$2$}, Animesh Mukherjee\textsuperscript{$2$}\\
	\textsuperscript{$1$} Indian Institute of Engineering Science and Technology, Shibpur\\	
	\textsuperscript{$2$} Indian Institute of Technology, Kharagpur\\
	{\tt \small punyajoysaha1998@gmail.com  , binnymathew@iitkgp.ac.in} \\
	{\tt \small pawang.iitk@gmail.com, animeshm@gmail.com}%
}

\date{}

\begin{document}
\maketitle
\begin{abstract}
  \textbf{English.} With the online proliferation of hate speech, there is an urgent need for systems that can detect such harmful content. In this paper, We present the machine learning models developed for the Automatic Misogyny Identification (AMI) shared task at EVALITA 2018. We generate three types of features: Sentence Embeddings, TF-IDF Vectors, and BOW Vectors to represent each tweet. These features are then concatenated and fed into the machine learning models. Our model came \textbf{First} for the English Subtask A and \textbf{Fifth} for the English Subtask B. We release our winning model for public use\footnote{\label{model_link}\url{https://github.com/punyajoy/Hateminers-EVALITA}}.
\end{abstract}

\begin{abstract-alt}
 \textrm{\bf{Italiano.}} Con la proliferazione online di incitamento all'odio, c'è un'urgente necessità di sistemi in grado di rilevare tali contenuti dannosi. In questo documento, presentiamo i modelli di apprendimento automatico sviluppati per l'attività di identificazione automatica Misogyny Identification (AMI) a EVALITA 2018. Generiamo tre tipi di funzionalità: Embedded di frasi, Vettori TF-IDF e Vettori BOW per rappresentare ciascun tweet. Queste caratteristiche vengono quindi concatenate e inserite nei modelli di apprendimento automatico. Il nostro modello è arrivato \ textbf {Primo} per la sottotabella inglese A e \ textbf {Fifth} per la sottotabella inglese B. Rilasciamo il nostro modello vincente per uso pubblico\footref{model_link}.
\end{abstract-alt}

\section{Introduction}
 Twitter defines hateful misconduct as ``\textit{you may not promote violence against or directly attack or threaten other people on the basis of race, ethnicity, national origin, sexual orientation, gender, gender identity, religious affiliation, age, disability, or serious disease}\footnote{\url{https://goo.gl/RSSrrZ}}''. With the online proliferation of hate speech, several countries like USA, Germany, and France have laws to ban such hateful content. This situation calls for online hate detection systems that are necessary to curb the rapidly increasing hate speech. In particular, there is a rise in online violence against women (Misogyny).

According to Pew research\footnote{\url{http://www.pewinternet.org/2017/07/11/online-harassment-2017}}, women encounter sexualized forms of abuse at much higher rates than men. Sites like Twitter are failing in acting promptly against online mysogyny and taking too much time to remove the content\footnote{\url{https://goo.gl/zbYTWA}}. The research community has now started to focus on this issue and is developing methods to detect online mysogyny~\cite{hewitt2016problem,fersini2018overview,poland2016haters}.

In this paper, we focus on detection of misogynous posts in Twitter that are written in English and describe our submission (\textit{Hateminers}) for the task of Automatic Misogyny Identification (AMI) at EVALITA2018~\cite{fersini2018overviewEvalita}. We concatenate three types of features to represent each tweet and use machine learning models for classification.

For the English Task  A,  we are ranked $1^{st}$ (team  ``Hateminers")  at  the  AMI shared task at EVALITA  2018 competition,  with  an  accuracy  of  70.4\%.  For  the English Task  B,  we are ranked $5^{th}$ (team rank $3^{rd}$), with an average macro-average F1-score of 0.37.

\section{Related works}

The research on hatespeech is gaining momentum with several works which focus on different aspects such as analyzing hatespeech~\cite{ElSherief2018PeerTP,mathew2018spread,silva2016analyzing,chandrasekharan2017you,grondahl2018all}, and detection of hatespeech~\cite{fortuna2018survey,davidson2017automated,qian2018hierarchical}. 

Recently, there seems to be growing interest in the identification of misogynous contents online~\cite{ging2018special}. Some of the initial works on identification of misogynous contents online were performed by ~\cite{Hewitt:2016:PIM:2908131.2908183}. 
In ~\cite{fox2015perpetuating}, the authors study the roles of anonymity and interactivity in response to sexist content posted on a social networking site. They concluded that interacting with sexist content anonymously promotes greater hostile sexism than interacting with it using an identified account.

\section{Dataset and task description}

The AMI shared task at EVALITA2018 had two balanced datasets for the English and Italian language. We participated in the English language shared task only. So, we present the systems developed for the English language AMI task only.

\subsection{Dataset}
The training dataset consisted of 4000 labelled tweets and the test dataset had 1000 unlabelled tweets. The distribution of different labels is presented in Table~\ref{tab:label_distribution}.
The English corpora have been manually labelled by several annotators according to three levels:
\begin{itemize}
    \item     Misogyny (Misogyny vs Not Misogyny)
    \item Misogynistic category (discredit, derailing, dominance, sexual harassment \& threats of violence, stereotype \& objectification)
    \item Target (active vs passive)
\end{itemize}

As observed from Table~\ref{tab:label_distribution}, the label distribution for Task A is balanced, while in Task B the
distribution is highly unbalanced for both misogyny behaviors and targets. We will explain these categories in the following section.

\begin{table}[!htbp]
	
\resizebox{\linewidth}{!}{\begin{tabular}{l l l l } 
\hline \hline
Type & Labels  & Training & Test \\ \hline
\multirow{2}{*}{\parbox{0.25\linewidth}{\centering Misogyny}}  &   Misogyny    & 1785  & 540   \\
&   Non-Misogyny    & 2215   & 460   \\ \hline
\multirow{5}{*}{\parbox{0.25\linewidth}{\centering Misogynistic category}}&   Discredit   & 1014   & 141   \\
&   Derailing   & 92  & 11  \\
&   Dominance   & 148  & 124  \\
&   Sexual Harassment    & 352  & 44  \\
&   Stereotype    & 179  & 140   \\ \hline
\multirow{2}{*}{\parbox{0.25\linewidth}{\centering Misogyny Target}}&   Active   & 1058   & 401   \\
&   Passive     & 727  & 59  \\
		
		\hline
	\end{tabular}}
	\caption{The distribution of different labels in the English language dataset.}
	~\label{tab:label_distribution}
\end{table}
\subsection{Tasks}

\noindent\textbf{Task A: }
First, it is asked to have a binary classification of the tweets, that is as either misogynous or not misogynous. The performance of the system is measured based on the accuracy.

\noindent\textbf{Task B: }
Next, it is asked to classify the misogynous tweets according to both the misogynistic behaviour and the target of the message. The evaluation metric is macro F1-score for this task.

A tweet must be classified uniquely within one of the following categories:

\begin{enumerate}
    \item \textbf{Stereotype \& objectification:} a widely held but fixed and oversimplified image or idea of a woman; description of women's physical appeal and/or comparisons to narrow standards.
    \item \textbf{Dominance:} to assert the superiority of men over women to highlight gender inequality.
    \item \textbf{Derailing:} to justify woman abuse, rejecting male responsibility; an attempt to disrupt the conversation in order to redirect  women's conversations on something more comfortable for men.
    \item \textbf{Sexual harassment \& threats of violence:} to describe actions as sexual advances, requests for sexual favors, harassment of a sexual nature; intent to physically assert power over women through threats of violence.
    \item \textbf{Discredit:} slurring over women with no other larger intention.
\end{enumerate}

On the other hand, the target classification is again binary:
\begin{enumerate}
    \item \textbf{Active (individual):} the text includes offensive messages purposely sent to a specific target.
    \item \textbf{Passive (generic):} it refers to messages posted to many potential receivers.
\end{enumerate}

\section{System description}
In this section, we will explain the details regarding the features and machine learning models used for the task.

\subsection{Feature generation}
\noindent\textbf{Pre-processing: }
We pre-process the tweets before performing the feature extraction. The following steps were followed:

\begin{itemize}
    \item We remove all the URLs.
    
    \item Convert tweet text to lowercase.
    
    \item Words such as ``ain't'', ``i'll'' were replaced by the corresponding expanded forms.
    
    \item Removed emojis, stop words, and punctuation.
    
    \item Performed tokenization and stemming.
\end{itemize}

\noindent\textbf{Feature vector: }
The pre-processed tweets were used to generate the features for the classifiers. We generated three types of features and concatenate them for each tweet. We experimented with all the features and found that the combination of all the features worked the best.  We explain each of the feature type below.

\begin{itemize}
    \item \textbf{Sentence embeddings:}
The sentence vector is generated using an Universal Sentence Encoder~\cite{cer2018universal} which outputs a 512 dimensional vector representation of the text. Recent works~\cite{conneau2017supervised} have shown stronger performance using pre-trained sentence level embeddings as compared to word level embeddings. We provide each of the preprocessed tweets as input to the sentence encoder and use the vector output for our task.

\item \textbf{TF-IDF vector: }
TF-IDF vectors were generated using Scikit's\footnote{\url{https://goo.gl/9FrZLD}} TF-IDF vectorizer on the pre-processed tweets.

\item \textbf{Bag of words vector (BoWV): }
The BoWV approach uses the average of the GloVe~\cite{pennington2014glove} word embeddings to represent a sentence. We set the size of the vector embeddings to 300.

\end{itemize}

\subsection{Classifiers used} 
We experiment with three machine learning models for Task A \& B.

\noindent\textbf{Logistic Regression (LR):} We use the LR implementation available in scikit-learn\footnote{\url{https://scikit-learn.org/stable/modules/generated/sklearn.linear_model.LogisticRegression.html}}. We set $C$ as 1.0 for all the tasks.

\noindent\textbf{XGBoost (XGB):} XGB\footnote{\url{https://github.com/dmlc/xgboost}} is an optimized distributed gradient boosting library designed to be highly efficient, flexible and portable. we set the $objective$ parameter as `binary:logistic' $scale\_pos\_weight$ was set to 0.8 and $reglamda$ set to 3.0.

\noindent\textbf{CatBoost (CB):}
CB~\cite{dorogush2017catboost} is a state-of-the-art open-source gradient boosting on decision trees library developed by Yandex\footnote{\url{https://catboost.ai}}. We set $scale\_pos\_weight$ to 0.8 for all experiments.




\begin{figure}[!t]
	\centering
	\includegraphics[width=.47\textwidth]{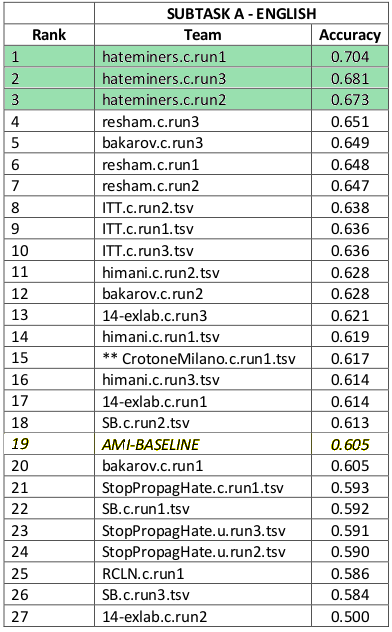}
	\caption{AMI results for the English Subtask A (Misogyny classification). Our System came at the top position.}
	\label{fig:english_subtask_A}
\end{figure}

\begin{figure*}[h!]
	\centering
	\includegraphics[width=.95\textwidth]{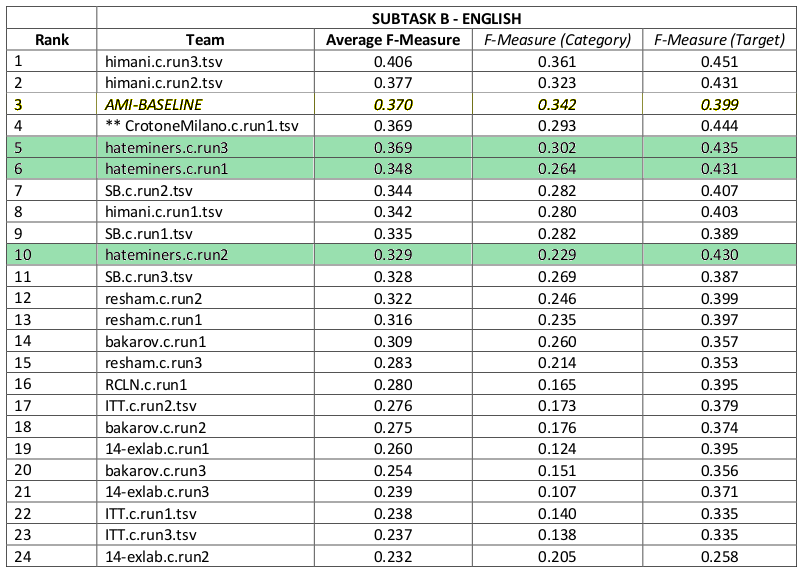}
	\caption{AMI results for the English Subtask B (Category and target classification). Out best system came at 5$^{th}$ position (3$^{rd}$ best team).}
	\label{fig:english_subtask_B}
\end{figure*}
\section{Results: }

The results of our system for English Task A are presented in Figure~\ref{fig:english_subtask_A} and for English Task B are presented in Figure~\ref{fig:english_subtask_B}. Our systems captured the top three ranks for the English Subtask A of Misogyny identification. For Subtask B, our best system came at fifth position ($3^{rd}$ best team).

We obtained the best result for English Subtask A in run\#1 (0.704 accuracy) in which we used Logistic Regression classifier and had the $1^{st}$ rank. Our other two runs, both using Catboost model, ranked $2^{nd}$ and $3^{rd}$ in the task.

For the English Subtask B, in which we needed to classify the category and target of misogyny, we kept the same set of features as we used for Task A. Our best system ranked $5^{th}$ (0.37 average F-Measure for run\#3) which used Catboost classifier for both category and target classification.

\section{Discussion}

We found that our system was able to achieve good performance for classifying the targets, but was not able to perform good in category classification. On closer inspection of Subtask B results\footnote{\url{https://amievalita2018.files.wordpress.com/2018/11/english-detailed-results-category-target.pdf}}, we found that the main reason for the poor performance was the high data imbalance. We observe that several of the submitted systems perform poorly on the different categories of Task B. The under represented categories such as \textit{DERAILING} and \textit{DOMINANCE} were hard to detect due to the data imbalance.

\section{Conclusion}

In this paper we present our approach to detect misogynous tweets in twitter. We generate sentence embeddings, TF-IDF Vectors, and BOW vectors for each tweet and then concatenate them. These vectors are then used as features for models such as CatBoost and Logistic Regression. Our model occupied the top three positions for English Subtask A and our best model for English Subtask B came at $5^{th}$ rank ($3^{rd}$ best team).


We have also made the winning model public\footref{model_link} for other researchers to use.

\bibliographystyle{acl}
\bibliography{Main}

\begin{thebibliography}{}

\bibitem[\protect\citename{Cer \bgroup et al.\egroup }2018]{cer2018universal}
Daniel Cer, Yinfei Yang, Sheng-yi Kong, Nan Hua, Nicole Limtiaco, Rhomni~St
  John, Noah Constant, Mario Guajardo-Cespedes, Steve Yuan, Chris Tar, et~al.
\newblock 2018.
\newblock Universal sentence encoder.
\newblock {\em arXiv preprint arXiv:1803.11175}.

\bibitem[\protect\citename{Chandrasekharan \bgroup et al.\egroup
  }2017]{chandrasekharan2017you}
Eshwar Chandrasekharan, Umashanthi Pavalanathan, Anirudh Srinivasan, Adam
  Glynn, Jacob Eisenstein, and Eric Gilbert.
\newblock 2017.
\newblock You can't stay here: The efficacy of reddit's 2015 ban examined
  through hate speech.
\newblock {\em Proceedings of the ACM on Human-Computer Interaction},
  1(CSCW):31.

\bibitem[\protect\citename{Conneau \bgroup et al.\egroup
  }2017]{conneau2017supervised}
Alexis Conneau, Douwe Kiela, Holger Schwenk, Lo{\"\i}c Barrault, and Antoine
  Bordes.
\newblock 2017.
\newblock Supervised learning of universal sentence representations from
  natural language inference data.
\newblock In {\em Proceedings of the 2017 Conference on Empirical Methods in
  Natural Language Processing}, pages 670--680.

\bibitem[\protect\citename{Davidson \bgroup et al.\egroup
  }2017]{davidson2017automated}
Thomas Davidson, Dana Warmsley, Michael Macy, and Ingmar Weber.
\newblock 2017.
\newblock Automated hate speech detection and the problem of offensive
  language.
\newblock {\em arXiv preprint arXiv:1703.04009}.

\bibitem[\protect\citename{Dorogush \bgroup et al.\egroup
  }2017]{dorogush2017catboost}
Anna~Veronika Dorogush, Vasily Ershov, and Andrey Gulin.
\newblock 2017.
\newblock Catboost: gradient boosting with categorical features support.

\bibitem[\protect\citename{ElSherief \bgroup et al.\egroup
  }2018]{ElSherief2018PeerTP}
Mai ElSherief, Shirin Nilizadeh, Dana Nguyen, Giovanni Vigna, and Elizabeth~M.
  Belding-Royer.
\newblock 2018.
\newblock Peer to peer hate: Hate speech instigators and their targets.
\newblock In {\em ICWSM}.

\bibitem[\protect\citename{Fersini \bgroup et al.\egroup
  }2018a]{fersini2018overviewEvalita}
Elisabetta Fersini, Debora Nozza, and Paolo Rosso.
\newblock 2018a.
\newblock Overview of the evalita 2018 task on automatic misogyny
  identification (ami).

\bibitem[\protect\citename{Fersini \bgroup et al.\egroup
  }2018b]{fersini2018overview}
Elisabetta Fersini, Paolo Rosso, and Maria Anzovino.
\newblock 2018b.
\newblock Overview of the task on automatic misogyny identification at ibereval
  2018.

\bibitem[\protect\citename{Fortuna and Nunes}2018]{fortuna2018survey}
Paula Fortuna and S{\'e}rgio Nunes.
\newblock 2018.
\newblock A survey on automatic detection of hate speech in text.
\newblock {\em ACM Computing Surveys (CSUR)}, 51(4):85.

\bibitem[\protect\citename{Fox \bgroup et al.\egroup
  }2015]{fox2015perpetuating}
Jesse Fox, Carlos Cruz, and Ji~Young Lee.
\newblock 2015.
\newblock Perpetuating online sexism offline: Anonymity, interactivity, and the
  effects of sexist hashtags on social media.
\newblock {\em Computers in Human Behavior}, 52:436--442.

\bibitem[\protect\citename{Ging and Siapera}2018]{ging2018special}
Debbie Ging and Eugenia Siapera.
\newblock 2018.
\newblock Special issue on online misogyny.

\bibitem[\protect\citename{Gr{\"o}ndahl \bgroup et al.\egroup
  }2018]{grondahl2018all}
Tommi Gr{\"o}ndahl, Luca Pajola, Mika Juuti, Mauro Conti, and N~Asokan.
\newblock 2018.
\newblock All you need is" love": Evading hate-speech detection.
\newblock {\em arXiv preprint arXiv:1808.09115}.

\bibitem[\protect\citename{Hewitt \bgroup et al.\egroup
  }2016a]{Hewitt:2016:PIM:2908131.2908183}
Sarah Hewitt, T.~Tiropanis, and C.~Bokhove.
\newblock 2016a.
\newblock The problem of identifying misogynist language on twitter (and other
  online social spaces).
\newblock In {\em Proceedings of the 8th ACM Conference on Web Science}, WebSci
  '16, pages 333--335. ACM.

\bibitem[\protect\citename{Hewitt \bgroup et al.\egroup
  }2016b]{hewitt2016problem}
Sarah Hewitt, Thanassis Tiropanis, and Christian Bokhove.
\newblock 2016b.
\newblock The problem of identifying misogynist language on twitter (and other
  online social spaces).
\newblock In {\em Proceedings of the 8th ACM Conference on Web Science}, pages
  333--335. ACM.

\bibitem[\protect\citename{Mathew \bgroup et al.\egroup
  }2018]{mathew2018spread}
Binny Mathew, Ritam Dutt, Pawan Goyal, and Animesh Mukherjee.
\newblock 2018.
\newblock Spread of hate speech in online social media.
\newblock {\em arXiv preprint arXiv:1812.01693}.

\bibitem[\protect\citename{Pennington \bgroup et al.\egroup
  }2014]{pennington2014glove}
Jeffrey Pennington, Richard Socher, and Christopher Manning.
\newblock 2014.
\newblock Glove: Global vectors for word representation.
\newblock In {\em EMNLP}, pages 1532--1543.

\bibitem[\protect\citename{Poland}2016]{poland2016haters}
Bailey Poland.
\newblock 2016.
\newblock {\em Haters: Harassment, abuse, and violence online}.
\newblock U of Nebraska Press.

\bibitem[\protect\citename{Qian \bgroup et al.\egroup
  }2018]{qian2018hierarchical}
Jing Qian, Mai ElSherief, Elizabeth Belding, and William~Yang Wang.
\newblock 2018.
\newblock Hierarchical cvae for fine-grained hate speech classification.
\newblock In {\em Proceedings of the 2018 Conference on Empirical Methods in
  Natural Language Processing}, pages 3550--3559.

\bibitem[\protect\citename{Silva \bgroup et al.\egroup
  }2016]{silva2016analyzing}
Leandro~Ara{\'u}jo Silva, Mainack Mondal, Denzil Correa, Fabr{\'\i}cio
  Benevenuto, and Ingmar Weber.
\newblock 2016.
\newblock Analyzing the targets of hate in online social media.
\newblock In {\em ICWSM}, pages 687--690.

\end{thebibliography}

\end{document}